\documentclass[a4paper]{jpconf}
\usepackage{graphicx}
\usepackage{amsmath}

\begin{document}
\title{Geometric Allocation Approaches in Markov Chain Monte Carlo}

\author{S Todo$^1$ and H Suwa$^2$}

\address{$^1$ Institute for Solid State Physics, University of Tokyo, 7-1-26-R501 Port Island South, Kobe 650-0047, Japan}
\address{$^2$ Department of Physics, Boston University, 590 Commonwealth Avenue, Boston, Massachusetts 02215, USA}

\ead{wistaria@issp.u-tokyo.ac.jp}

\begin{abstract}
The Markov chain Monte Carlo method is a versatile tool in
statistical physics to evaluate multi-dimensional integrals
numerically.  For the method to work effectively, we must consider the
following key issues: the choice of ensemble, the selection of
candidate states, the optimization of transition kernel, algorithm for
choosing a configuration according to the transition probabilities.  We show that the
unconventional approaches based on the geometric allocation of
probabilities or weights can improve the dynamics and scaling of the
Monte Carlo simulation in several aspects.  Particularly, the approach using the
irreversible kernel can reduce or sometimes completely eliminate the
rejection of trial move in the Markov chain.  We also discuss how the space-time
interchange technique together with Walker's method of aliases
can reduce the computational time especially for
the case where the number of candidates is large, such as models with
long-range interactions.
\end{abstract}

\section{Introduction}

The Monte Carlo method for high-dimensional problems has a wide
variety of applications as a versatile and interdisciplinary
computational tool~\cite{RobertC2004}.  There are many interesting
topics on strongly correlated systems in physics, e.g., the phase
transitions and novel exotic phases, where the dimension of the state
space for capturing the essential physics is often more than hundreds
of thousand. The Markov chain Monte Carlo (MCMC) method that overcomes
the curse of dimensionality has been effectively applied to many
problems in high dimensions~\cite{LandauB2005}.

In principle, the MCMC method achieves a statistically exact importance
sampling from any probability distribution function by introducing a special
kind of random walk in the state space.  The MCMC
method keeps the sampling efficiency reasonably high even in
high-dimensional problems.
Instead of the curse of dimensionality, however,
the MCMC method often suffers from the sample correlation. Since the next
configuration is generated (updated) from the previous one, the
samples are not independent of each other. Then the correlation gives
rise to two problems; we have to wait for the distribution convergence
(equilibration) before sampling, and the number of effective samples
is decreased. The former convergence problem is quantified by the
distance to the target distribution~\cite{Tierney1994} or the spectral
gap of the transition kernel. As an assessment for the latter problem,
the decrease of the number of effective samples, the integrated
autocorrelation time is defined as
\begin{eqnarray}
  \tau_{\rm int} = \sum_{t=1}^{\infty} C(t) = \sum_{t=1}^{\infty}
  \frac{\langle O_{i+t}O_i \rangle - \langle O \rangle^2}{\langle O^2
    \rangle - \langle O \rangle^2},
\end{eqnarray}
where $O_i$ is an observable at the $i$-th Monte Carlo step, and
$C(t)$ is ideally independent of $i$ after the distribution
convergence.  In terms of the autocorrelation time, the number of
effective samples is roughly given by
\begin{eqnarray}
M_{\rm eff} \simeq \frac{M}{1 + 2 \tau_{\rm int}},
\end{eqnarray}
where $M$ is the total number of samples, i.e., the total Monte Carlo
steps, in the simulation.  Although an MCMC method satisfying the
appropriate conditions, the balance condition and the ergodicity, guarantees asymptotically correct results in
principle~\cite{MeynT1993}, variance reduction of relevant estimators
is crucial for the method to work in practice. If the central limit
theorem holds, the variance of expectations decreases as $v/M \simeq
{\rm var}(f) / M_{\rm eff}$, where $v$ is called the asymptotic
variance that depends on the integrand function and the update method
through the autocorrelation time.

In general, we should take the following four key points into account
for achieving efficient updates in the MCMC method:
\begin{enumerate}
\item Choice of ensemble and assignment (or modification)
  of the weights of configurations.
\item Generation of set of candidate configurations from the
  current configuration.
\item Construction of the transition kernel (probability), given a
  set of candidate configurations.
\item Algorithm for choosing a configuration among the candidates
  according to the transition probability.
\end{enumerate}
As for (i), a series of extended ensemble methods, such as the
multicanonical method~\cite{BergN1992}, Wang-Landau
method~\cite{WangL2001a}, simulated tempering~\cite{MarinariP1992},
and exchange Monte Carlo method~\cite{HukushimaN1996}, have been
proposed and successfully applied to the protein folding problems, the
spin glasses, etc.  The cluster algorithms, e.g., the Swendsen-Wang
algorithm~\cite{SwendsenW1987} and the loop
algorithm~\cite{EvertzLM1993}, which overcome the critical slowing
down near the critical point (the correlation-time growth in a power-law form of the system
size) by taking advantage of mapping to graph
configurations~\cite{Todo2013a}, are the representative improvements
from the standpoint of (ii).  The hybrid (Hamiltonian) Monte Carlo
method~\cite{DuaneKPR1987} that performs a simultaneous move by
generating the candidate state by a Newtonian dynamics with an
artificial momentum is another successful example of (ii).

On the other hand, the key points (iii) and (iv) in the above list
have {\em not} been studied in great depth so far.  In this paper, we
propose new approaches that are based on geometrical procedure on
probabilities or weights.  As we will show below, our unconventional
approaches can improve the dynamics and scaling of the Monte Carlo
simulation from the standpoints (iii) and (iv).

\section{Markov chain Monte Carlo and balance condition}

In the MCMC method, the {\em ergodicity} (irreducibility) of the Markov chain
guarantees the consistency of estimators; the Monte Carlo average
asymptotically converges in probability to a unique value irrespective
of the initial configuration.  The {\em total balance}, the
invariance of the target distribution, is usually imposed though
some interesting adaptive procedures have been proposed these
days. Since the invention of the MCMC method in
1953~\cite{MetropolisRRTT1953}, the reversibility, namely the {\em detailed
balance}, has been additionally imposed in most practical simulations
as a sufficient condition for the total balance.  Under the detailed
balance, every elementary transition is forced to balance with a
corresponding inverse process.  Thanks to this condition, it becomes
practically easy to find qualified transition probabilities in actual
simulations. The standard update methods, such as the Metropolis
(Metropolis-Hastings) algorithm~\cite{MetropolisRRTT1953,Hastings1970}
and the heat bath algorithm (Gibbs sampler)~\cite{GemanG1984}, satisfy
the reversibility. The performance of these seminal update methods has
been analytically and numerically investigated in many papers.

There is a simple theorem about the reversible kernel as a guideline
for the optimization of the transition matrix. Now we
define an order of the matrices as $P_2 \geq
P_1$ for any two transition matrices if each of the off-diagonal
elements of $P_2$ is grater than or equal to the corresponding
off-diagonal elements of $P_1$. The following statement is Theorem
2.1.1 of \cite{Peskun1973}.

\begin{quote}
{\bf [Peskun (1973)]} 
Suppose each of the irreducible transition matrices $P_1$ and $P_2$ is
reversible for a same invariant probability distribution $\pi$. If
$P_2 \geq P_1$, then, for any $f$,
\begin{equation}
v(f, P_1, \pi ) \geq v( f, P_2, \pi ),
\end{equation}
where
\begin{equation}
v(f,P,\pi) = \lim_{M \rightarrow \infty} M\, {\rm var}( \hat{I}_M ),
\end{equation}
and $\hat{I}_M = \sum_{i=1}^M f(x_i)/M$ is an estimator of $I =
E_{\pi}(f)$ using $M$ samples, $x_1, x_2, ..., x_M$, of the Markov
chain generated by $P$.
\label{theo:p}
\end{quote}
According to this theorem, a modified Gibbs sampler, called
``Metropolized Gibbs sampler,'' was
proposed~\cite{FrigessiHY1992,Liu1996a}.  By the usual Gibbs sampler,
we choose the next state with forgetting the current state. By the
Metropolized version, on the other hand, a candidate is chosen except
the current state and it will be accepted/rejected by using the
Metropolis scheme.
This modified Gibbs sampler is reduced not to the usual Gibbs sampler
but to the Metropolis algorithm in the case where the number of
candidates is two.  It is proved that the modified Gibbs sampler has a
smaller asymptotic variance $v$ than the original
one~\cite{FrigessiHY1992,Liu1996a}.
Furthermore, an iterative version of the Metropolized Gibbs sampler
was proposed as well~\cite{FrigessiHY1992}. What we have learned from
the Peskun's theorem is the guideline that the rejection rate (the
diagonal elements of the transition matrix) should be minimized in
general.

As we see, most of optimizations of the transition matrix have
been proposed within the detailed balance.
However, the reversibility is not a necessary condition for the
invariance of the target distribution.  The sequential update, where
the state variables are swept in a fixed order, breaks the detailed
balance, but can satisfy the total
balance~\cite{ManousiouthakisD1999}. In the meanwhile, some
modifications of reversible chain into an irreversible chain have been
proposed so far.  One example is a method of the duplication of the
state space with an additional variable, such as a direction on an
axis~\cite{DiaconisHN2000,TuritsynCV2011,SakaiH2013}, and it has been
shown that the autocorrelation time can be reduced drastically at
least for specific models.  The axis can be a combination of the state
variables, the energy, or any quantity. The extended version with the
multi-axes has been applied to some physical
models~\cite{FernandesW2011}.  Also the artificial momentum in the
hybrid Monte Carlo method~\cite{DuaneKPR1987} performs partly as a
direction in the state space.  A similar idea with the addition of a
direction has been proposed~\cite{Neal2004}, where the next state in
the Markov chain is generated depending on not only one step before
but also the two (several) steps before.  Then the resulting Markov
chain can be irreversible because the history of the states has the
direction. As other approaches, inserting a probability vortex in the
state space was discussed~\cite{SunGS2010}, an asymmetric choice of
the heading direction was applied in a hard-sphere
system~\cite{BernardKW2009}, and a global optimization of the
transition matrix was discussed~\cite{HwangCCP2012}. As seen above,
the role of the net stochastic flow (irreversible drift) has caught
the attention~\cite{HwangHS2005}.  Note that the hybrid Monte Carlo
method seems to break the detailed balance in the extended state
space, but it is not essential because the additional update of the
artificial momentum easily recovers the reversibility.

So far, most of the irreversible chains were based on the reversible
update methods, such as the Metropolis algorithm.  A more significant
breaking of the reversibility can be achieved by applying the methods
we will explain in this paper.  In the following we introduce a new
type of method breaking the detailed balance~\cite{SuwaT2010}, which
applies a geometric approach to solve the algebraic equations.

\section{Geometric construction of irreversible kernel}

In the MCMC method for a lattice system, the configuration is locally
updated, and the huge transition kernel or matrix is implicitly
constructed by the consecutive local updates. Following 
\cite{SuwaT2010}, let us consider a local update of a discrete
variable as an elementary process. Now we have $n$ next candidate
configurations including the current one. The weight of each
configuration is given by $w_i$ $(i=1,...,n)$, to which the target
probability measure $\pi_i$ is in proportion. We introduce a quantity
$v_{ij}=w_i p_{i \rightarrow j}$ that corresponds to the amount of
(raw) stochastic flow from state $i$ to $j$. The law of probability
conservation and the total balance are expressed as
\begin{eqnarray}
w_i &=& \sum_{j=1}^n v_{ij} \qquad \forall i \label{eqn:cl}\\
w_j &=& \sum_{i=1}^n v_{ij} \qquad \forall j \label{eqn:bc},
\end{eqnarray}
respectively. The average rejection rate is written as $\sum_i v_{ii}
/ \sum_i w_i $. It is easily confirmed that the Metropolis algorithm
with the flat proposal distribution gives
\begin{equation}
v_{ij} = \frac{1}{n-1} \min[ w_i, w_j] \qquad i \neq j,
\end{equation}
and the heat bath algorithm (Gibbs sampler) does
\begin{equation}
v_{ij} = \frac{ w_i w_j }{ \sum_{k=1}^n w_k } \qquad \forall i, j.
\end{equation}
The both satisfy the above conditions~(\ref{eqn:cl})
and~(\ref{eqn:bc}).  The reversibility is manifested by the symmetry
under the interchange of the indices:
\begin{equation}
v_{ij} = v_{ji} \qquad \forall i,j.
\label{eqn:dbc}
\end{equation}

\begin{figure}[t]
\includegraphics[width=20.2pc]{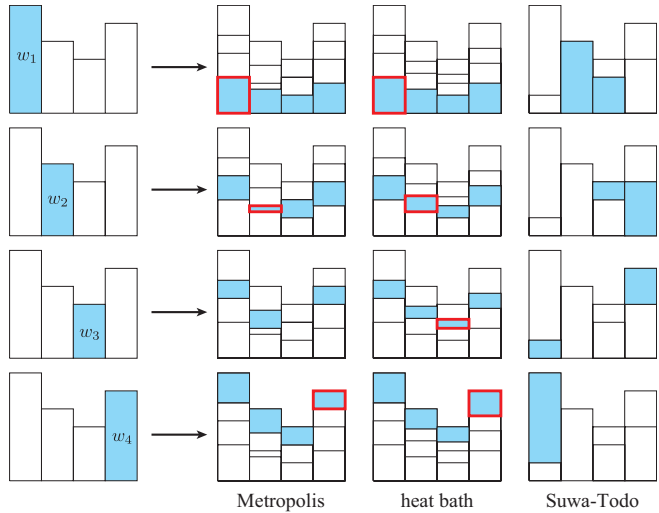} \hfill 
\begin{minipage}[b]{15.2pc}\caption{Example of weight landfill by the
 Metropolis, heat bath, and Suwa-Todo algorithms for $n=4$.  Our
 algorithm is rejection free, while there remain finite rejection rates
 in the conventional methods as indicated by the red thick frames (from 
\cite{SuwaT2010}).}\label{fig:landfill} 
\end{minipage}
\end{figure}

The aim here is to find a set $\{ v_{ij}\}$ that minimizes the average
rejection rate (the diagonal elements of the transition matrix) under
the conditions~(\ref{eqn:cl}) and~(\ref{eqn:bc}).  The procedure for
the task can be understood visually as {\it weight allocation}, where
we move (or allocate) weight $(v_{ij})$ from $i$-th state to $j$-th
box with the entire shape of the weight boxes kept intact
(figure~\ref{fig:landfill}). We propose the following
algorithm~\cite{SuwaT2010}:
\begin{enumerate}
\item Choose a configuration with maximum weight among the
  candidates. If two or more configurations have the same maximum
  weight, choose one of them. In the following, we assume $w_1$ is the
  maximum without loss of generality. The order of the remaining
  weights does not matter.
\item Allocate the maximum weight $w_1$ to the next box ($i = 2$). If
  the weight still remains after saturating the box, reallocate the
  remainder to the next ($i = 3$). Continue until the weight is all
  allocated.
\item Allocate the weight of the first filled box ($w_2$) to the
  last partially filled box in step (ii). Continue the allocation
  likewise.
\item Repeat step (iii) for $w_3,w_4,\ldots w_n$. Once all the boxes
  with $i \ge 2$ are saturated, landfill the first box ($i = 1$)
  afterward.
\end{enumerate}
By this procedure (the index 1 is such
that $w_1$ has the maximum weight), $\{ v_{i \rightarrow j} \}$ are
determined as
\begin{equation}
v_{i \rightarrow j } = \max( 0, \, \min( \Delta_{ij}, \, w_i + w_j - \Delta_{ij}, \, w_i, \, w_j ) ),
\label{eqn:irreversible}
\end{equation}
where
\begin{eqnarray}
\Delta_{ij} &:=& S_i - S_{j-1} + w_1 \qquad 1 \leq i, \, j \, \leq n \\
S_i &:=& \sum_{k=1}^i w_k \qquad 1 \leq i \leq n \\
S_0 &:=& S_n.
\end{eqnarray}
It satisfies the conditions~(\ref{eqn:cl}) and~(\ref{eqn:bc}), but
breaks the reversibility; for example, $v_{12} >0 $ but $v_{21}=0$
in figure~\ref{fig:landfill}. The self-allocated weight that
corresponds the rejection rate is expressed as
\begin{eqnarray}
v_{ii} = \left\{ \begin{array}{lll} \displaystyle \max( 0, \, w_1 - \sum_{i=2}^n w_i ) & \qquad i = 1  \\
    0 \hspace{2mm} & \qquad i \geq 2.
  \end{array} \right.
\label{eqn:rejection}
\end{eqnarray}
That is, a rejection-free solution is obtained if the condition
\begin{eqnarray}
w_1 \leq \frac{S_n}{2} \equiv \frac{1}{2}\sum_{k=1}^n w_k
\label{eqn:rejection-free}
\end{eqnarray}
is satisfied.  When it is not satisfied, the maximum weight has to be
assigned to itself since it is larger than the sum of the rest.  Thus,
this solution is optimal in the sense that it minimizes the average
rejection rate. Furthermore, the rejection rate
expression~(\ref{eqn:rejection}) provides us a clear prospect; the
rejection rate is certainly reduced as the number of candidates is
increased. This idea was used to a quantum physical model and the
rejection rate was indeed reduced to zero~\cite{SuwaT2010}, and
extended to the case with continuous variables~\cite{Suwa2012}.

\begin{figure}[t]
\includegraphics[width=24pc]{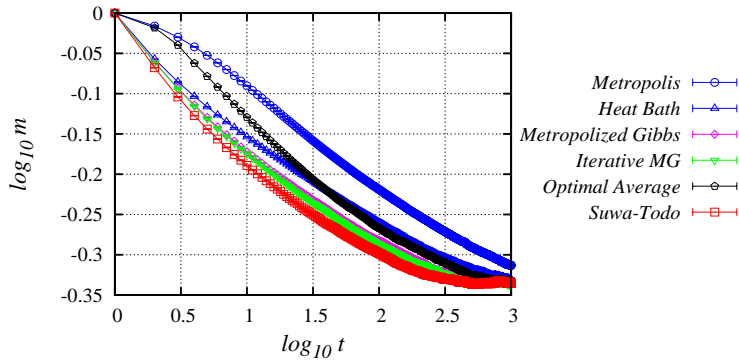} \hfill 
\begin{minipage}[b]{13.0pc}\caption{Convergence of magnetization squared
    in the ferromagnetic $4$-state Potts model on the square lattice
    with $L=32$ at the critical temperature. The horizontal axis is
    the Monte Carlo step. The simulation starts with the ordered (all
    ``up'') state.}\vspace*{.4em} \label{fig:relax}
\end{minipage}
\end{figure}

\begin{figure}[t]
\includegraphics[width=24pc]{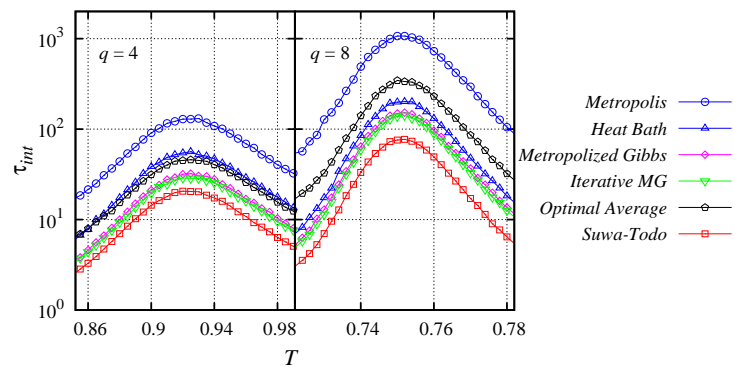} \hfill 
\begin{minipage}[b]{13pc}\caption{Autocorrelation time of order parameter near
  the transition temperature in the 4-state (left) and 8-state (right)
  Potts models. The system size is $16 \times 16$.}\vspace*{.4em}
   \label{fig:correlation}
\end{minipage}
\end{figure}

In Figs.~\ref{fig:relax} and \ref{fig:correlation}, the results of the
benchmark test of our new method are presented.  We investigate the
convergence (equilibration) and the autocorrelations in the
ferromagnetic $q$-state Potts model on the square
lattice~\cite{Wu1982}; the local state at site $k$ is
expressed as $\sigma_k$ that takes an integer ($1 \leq \sigma_k \leq
q$) and the energy is expressed as $H = -
\sum_{\langle i, j \rangle} \delta_{\sigma_i \sigma_j}$, where
$\langle i, j \rangle$ is a pair of sites connected on
the lattice. This system exhibits a continuous ($ q \leq 4$)
or first-order ($q > 4 $) phase transition at temperature $T=1 / \ln ( 1 +
\sqrt{q})$.  We calculate the square of order
parameter~\cite{Zheng1998} (structure factor) for $q=4$ and 8 by using the
several algorithms.

The order parameter convergence is shown in figure~\ref{fig:relax},
where the simulation starts with a fully-ordered (all ``up'') state
and the local variables are sequentially updated.  The square lattice
with $L=32$ on the periodic boundary conditions and the critical
temperature $T=1/\ln 3 = 0.9102392266\cdots$ are used.  The Metropolis algorithm, the
heat bath algorithm (Gibbs sampler), the Metropolized Gibbs
sampler~\cite{Liu1996a}, the iterative Metropolized Gibbs
sampler~\cite{FrigessiHY1992}, the optimal average
sampler~\cite{HwangCCP2012}, and the present algorithm
(Suwa-Todo)~\cite{SuwaT2010} are compared.  The validity of the all
update methods are confirmed by comparing the asymptotic estimator
convergence with each other (the Markov chain by the Gibbs sampler is
ergodic).  Our algorithm accomplishes the fastest convergence.  This
acceleration implies that the locally rejection-minimized algorithm
reduces the second largest eigenvalue of the whole transition matrix
in absolute value and increases the spectral gap of the Markov chain.

From figure~\ref{fig:correlation}, it is clearly seen that the present
algorithm also reduces the autocorrelation time significantly
in comparison with the conventional methods.  The
autocorrelation time $\tau_{\rm int}$ is estimated through the
relation: $ \sigma^2 \simeq ( 1 + 2 \tau_{\rm int} ) \sigma_{0}^2$,
where $\sigma_0^2$ and $\sigma^2$ are the variances of the estimator
without considering autocorrelation and with calculating correlation
from the binned data using a bin size much larger than the $\tau_{\rm
  int}$~\cite{LandauB2005}, respectively.  In the 4 (8)-state Potts
model, the autocorrelation time becomes nearly 6.4 (14) times as short
as that by the Metropolis algorithm, 2.7 (2.6) times as short as the
heat bath algorithm, and even 1.4 (1.8) times as short as the
iterative Metropolized Gibbs sampler. We investigated also the
dynamical exponent of the autocorrelation time at the critical
temperature.  Unfortunately, the locally optimized method does not
reduce the exponent.  The factor over 6, however, is always gained
against the Metropolis algorithm for all system sizes.

\section{Walker's method of aliases and its generalization}

As demonstrated in the previous section, increasing the number of
candidate configurations can generally reduce the rejection rate.
Then, we have to consider an efficient method to stochastically
generate an event among a number of candidates.
Such an event generation matters to various situations.
In the Swendsen-Wang cluster algorithm for the long-range
interacting spin models~\cite{LuijtenB1995}, for example, one has to choose $O(N)$
bonds among $O(N^2)$ candidates stochastically for making clusters, where $N$
is the number of sites (spins).  When the number of candidates
increases, the computational cost for choosing configurations may
dominate the total computing time.  As seen below, a method based on
geometric allocation of weights, called ``Walker's method of
aliases''~\cite{Walker1977,Knuth1997a}, and its
generalization~\cite{FukuiT2009} give an efficient solution for the
present problems.

Let us consider a random variable $X$ that takes an integral value
$x$ according to probabilities $\{P(x)\}$ ($1 \le x \le M$ and $\sum_x
P(x)=1$).  One of the simplest methods for generating such random
numbers is the one based on the rejection.  In this method the number
of iterations required until a random number is obtained is
approximately $M \times \max[P(x)]$ on average.  It means that the
method becomes inefficient quite rapidly as the variance of $P(x)$
increases.  Instead, the binary search on the table of cumulative
probabilities has been employed widely so far~\cite{LuijtenB1995},
where the number of operations can be reduced down to $O(\log M)$.

However, it is known that there exists a further effective method,
Walker's method of aliases, which is rejection free and generates a
random integer in a constant time.  The Walker algorithm requires two
tables of size $M$. One is the table of the modified probabilities $\{
C(x) \}$ ($0 \le C(x) \le 1$ $\forall x$) and the other is of integral
alias numbers $\{A(x)\}$ ($1 \le A(x) \le M$ $\forall x$).  One has to
set up these two tables in advance, so that
\begin{equation}
  P(x) = \frac{1}{M} \Big\{ C(x) + \sum_{k=1}^{M} [1-C(k)] \, \delta_{A(k),x} \Big\} \qquad \forall x
  \label{eqn:walker}
\end{equation}
is satisfied.  Using these $\{C(x)\}$ and $\{A(x)\}$, a random integer
is generated by the following procedure:
\begin{enumerate}
\item Generate a uniform integral random variable $r$ ($1 \le r \le M$).
\item Generate a uniform real random variable $u$ ($0 < u \le 1$).
\item If $u < C(r)$, then $x = r$.  Otherwise $x = A(r)$.
\end{enumerate}
This procedure includes no iterations, and thus completes in a constant time.

\begin{figure}[t]
\includegraphics[width=19.4pc]{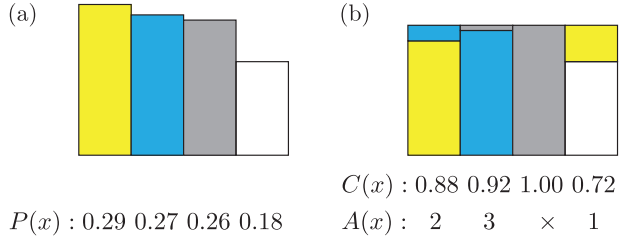} \hfill 
\begin{minipage}[b]{16.6pc}\caption{Example of (a) original probabilities $\{P(x)\}$, and (b) modified probabilities $\{C(x)\}$ and aliases $\{A(x)\}$ in Walker's method of aliases.  The area of each colour is the same in (a) and (b).} \label{fig:walker}
\end{minipage}
\end{figure}

The condition~(\ref{eqn:walker}) for the modified probabilities and
aliases can be illustrated graphically as demonstrated in
figure~\ref{fig:walker}.  In figure~\ref{fig:walker}(b), we have $M$ bins
of the same length.  Each bin contains at most two colours, the
original one and the colour pointed by its alias, and the total area of
each colour is the same as the original probability
(\ref{eqn:walker}).  Note that the existence of Walker's tables
for an arbitrary set of probabilities $\{P(x)\}$ is not trivial at
all.  However, a concrete and finite-step landfilling procedure of
geometric allocation has been given explicitly~\cite{FukuiT2009}, by
which the existence of the solution is guaranteed, albeit the
allocation is not unique and in general there exist many different
solutions.

This Walker's technique can be generalized to the following problem:
Let us consider $M$ bits each of which takes either 0 (deactivated) or
1 (activated) independently according to the probability $P(x)$ ($1
\le x \le M$ and $0 \le P(x) \le 1$ $\forall x$).  If one tries to
determine the state of each bit one by one, it will naturally take the time
proportional to $M$.  However, in the case where $\sum_x P(x)$ is much
smaller than $M$, i.e. most of $P(x)$'s are very small or vanishing,
one can adopt a different strategy by which the total computation time can
be reduced from $O(M)$ to $O(\sum_x P(x))$~\cite{FukuiT2009}.

Our central idea is assigning a nonnegative integer to each bit
instead of a binary, where zero (positive integers) corresponds to
deactivated (activated) state.  The integer to be assigned is
generated according to the Poisson distribution:
\begin{equation}
  f(\ell_x;\lambda_x) = \frac{e^{-\lambda_x} \lambda_x^{\ell_x}}{\ell_x!},
\end{equation}
where $\lambda_x$ is the mean of the distribution.  If one chooses
$\lambda_x$ as $-\log [1-P(x)]$, then $\ell_x$ takes a positive
integer with probability $P(x)$, since $\sum_{\ell_x \ge 1}
f(\ell_x;\lambda_x) = 1-e^{-\lambda_x}$.  At first glance, one might
think that the situation is getting worse by this modification, since
a Poisson random number, instead of a binary, is needed for determining
the state of each bit.  At this point, however, we leverage an
important property of the Poisson distribution; the Poisson process is
that for random events and there is no statistical correlation between
each two events.  It allows us to interchange the space and time axes, i.e.,
we can realize the whole distribution by
calculating just one Poisson random variable $\ell$, the total number
of events, from the distribution with mean $\lambda_{\rm tot} = \sum_x
\lambda_x$, and assigning each event to a bit afterward.  The
probability of choosing bit $x$ at least once is given by
\begin{equation}
  \begin{split}
  P_{\rm on}(x) &= \sum_{\ell=1}^\infty f(\ell;\lambda_{\rm tot})
  \sum_{k=1}^\ell \frac{\ell!}{(\ell-k)!k!}
  \Big(\frac{\lambda_x}{\lambda_{\rm tot}}\Big)^k
  \Big(1-\frac{\lambda_x}{\lambda_{\rm tot}}\Big)^{\ell-k} \\ &= 1 -
  \sum_{\ell=0}^\infty f(\ell;\lambda_{\rm tot})
  \Big(1-\frac{\lambda_x}{\lambda_{\rm tot}}\Big)^{\ell} = P(x).
  \end{split}
\end{equation}
By employing Walker's algorithm, this procedure completes in
$O(\lambda_{\rm tot})$ on average, which can be much smaller than $M$,
the number of bins, when most of $P(x)$'s are very small or
vanishing~\cite{FukuiT2009}.

As an example, let us consider the Swendsen-Wang cluster algorithm for
the $d$-dimensional ferromagnetic Ising model of $N$ sites with long-range
interactions, $J_{i,j} \sim r_{i,j}^{-\sigma}$, where $r_{i,j}$
denotes the distance between two lattice sites, $i$ and
$j$~\cite{LuijtenB1995}.  The exponent $\sigma$ should be larger than
$d$, otherwise the ground state energy becomes non-extensive.  At the
cluster formation procedure in the Swendsen-Wang
algorithm~\cite{SwendsenW1987}, each interacting bond is activated
with probability $P_{i,j} = 1-\exp(-2\beta J_{i,j})$, where $\beta$ is
the inverse temperature, and thus the total cost of this procedure is
proportional to $N^2$ in a naive implementation.  Since the
interaction becomes weaker as the distance increases, however, one may reduce
the computational cost by using the technique introduced above.
Indeed, in the present specific case
\begin{equation}
  \lambda_{\rm tot} = - \sum_{i,j} \log (1-P_{i,j}) = \sum_{i,j} 2\beta J_{i,j}
\end{equation}
which is of $O(N)$ as long as $\sigma > d$.  Thus, by using our 
  space-time interchange trick explained above,
the order of computational cost can be reduced greatly from $O(N^2)$
to $O(N)$~\cite{FukuiT2009} without any approximation, such as the
introduction of a finite cutoff.  One can extend this $O(N)$ technique
in various ways; the measurement of energy and specific heat,
Wang-Landau method~\cite{WangL2001a}, and exchange Monte
Carlo~\cite{HukushimaN1996}, as well as the cluster update for
long-range quantum spin models~\cite{FukuiT2009} and local-flip Monte Carlo update for
long-range frustrated spin models~\cite{Tomita2009}.

\section{Conclusion and discussions}

In the present paper, we have shown that the unconventional approaches based
on geometric allocation of probabilities or weights can improve the
dynamics and scaling of the Monte Carlo simulation in several aspects.
The approach using the irreversible kernel can reduce or sometimes
completely avoid the rejection of trial move in the Markov chain.  It
can be applied to any MCMC sampling including the system with continuous
variables and it is expected to improve the efficiency in general.
(For recent application to biomolecule systems, see e.g., \cite{ItohO2013,KondoT2013}.)
Also, the space-time interchange technique together with Walker's method
of aliases can reduce the computational
time especially for the case where the number of candidates is large.
We showed that by using this technique, the MCMC simulation for the
models with long-range interactions can be executed quite efficiently
in a rigorous way.  This technique is also generic and expected
to be applied to many MCMC methods.  The algorithm for
constructing an irreversible kernel for a generic set of weights has
been implemented in C++ and released as an open-source
library, BCL (Balance Condition Library)~\cite{BCLweb} together with
an efficient implementation of Walker's method of aliases.

The most simulations in the present paper has been done by using the
facility of the Supercomputer Center, Institute for Solid State
Physics, University of Tokyo. The simulation code has been developed
based on the ALPS library~\cite{ALPS2011,ALPSweb}.  The authors acknowledge the
support by the Grant-in-Aid for Scientific Research Program
(No.\,23540438) from JSPS, HPCI Strategic Programs for Innovative
Research (SPIRE) from MEXT, Japan, and the Computational Materials
Science Initiative (CMSI). HS is supported by the JSPS Postdoctoral
Fellow for Research Abroad.

\section*{References}

\bibliography{main}

\end{document}